\documentclass[preprintnumbers,amsmath,amssymb,onecolumn]{revtex4}
\usepackage[dvips,colorlinks=true,citecolor=red,filecolor=green,linkcolor=blue,pdfnewwindow=true]{hyperref}
\usepackage{graphicx}
\usepackage{color}
\usepackage{multirow}
\usepackage[title]{appendix}

\begin{document}
\title{Observation of complex functional cortical patterns in brain cognition}
\author{Jasleen Gund$^{1,2}$ and R.K. Brojen Singh$^1$}
\email{brojen@jnu.ac.in (Corresponding author)}
\affiliation{$^1$School of Computational and Integrative Sciences, Jawaharlal Nehru University, New Delhi-110067, India.\\$^{2}$Cognitive Brain Dynamics Lab, National Brain Research Centre, Manesar-122050, India.}

\begin{abstract}
{\noindent}Synaptic plasticity and neuron cross-talk are some of the important key mechanisms underlying formation of dynamic clusters of active neurons. The essence of this study is to model and decipher the mechanism of emergence of a task-specific functional connectivity among a population of neurons. We have used our proposed neuron activity pattern (NAP) model to define an assorted range of interactions and simulated across a 2D lattice with two-state (firing/non-firing) neurons. We observed the emergence of scale-free, hierarchically organized ordered patterns of active neurons, that we call as functional cortical patterns (FCPs), only at the near-critical phase transition. We have done extensive topological characterization of FCPs and tested its congruency with the functional brain networks obtained from the empirical electrophysiological data of a visual stimulus task. Our results of network theory attributes, fractal analysis and functional cartography have confirmed the transitioning effect while generation of FCPs. This gives us strong implication that the functional neuronal system supports far-from-equilibrium dynamics upon receiving a stimulus. Our investigation for the long, critical and short-ranged interactions has lead to an interpretation that near-critical transition phase is a spectrum of critical temperatures (coupling strengths) being affected by the distance between the neurons. The range of interaction among neurons plays an important role in defining the functional severity of the neuronal circuitry. We propose that interaction range as a function of coupling strength drives connectivity in an augmenting fashion as we observed an increase in coupling strength from short to long-range interactions, among the population of neurons. This gives us insight towards the intensity of cognitive behaviour, which is attained by multiple stimulus attempts.\\

\end{abstract}
 

\maketitle
\vskip 1cm
\section{Introduction}

{\noindent}The phenomenon of global features of complex cognitive functions in the brain arise from the organization of basic functions systematically coordinated in localized brain regions \citep{luria1980}. These two aspects of brain organization are reflected in brain cortical functions and evolve continuously through internal and external perturbations. Hence, human cognition can be thought of as a water surface where patterns in the form of ripples dwell on its surface upon receiving perturbations from outside. The fluidity of consciousness is reflected through the highly dynamic behaviour of brain activity \citep{Chialvo2010}. We can merely access those spatio-temporal patterns of brain activity through various electro-physiological and neuro-imaging techniques in order to understand the underlying dynamics. Electroencephalogram (EEG) is one such technique to measure electrical brain activity with high temporal resolution. The functional spatio-temporal clusters of neurons in the cortical layers of the brain is a consequence of various biological factors affecting the concentration of neurotransmitter, transfusion of ions through ion channels and voltage generated across the membrane of pre and post-synaptic neurons \citep{Klinshov2014, Voglis2006}. These biological factors determine the strength of the diffused signal among neurons and neuronal organization. Such recurring signals strengthen the circuitry of the neuron cluster allowing it to learn and memorize a specific function in the brain and rewiring among these complex functional patterns lead to varied cognitive abilities of the brain \citep{Colicos2006,Eguiluz2005a,Sporns2000,Sporns2002,Sporns2002a}. But, how the synchronous neuronal activation and generation of functional patterns lead to the execution of a cognitive task is a riddle to solve. The basis of understanding the complex cognitive functions would begin through studying the network dynamics of the interacting local and global neurons. In shaping the functional network topology of the brain, the interaction range of the cortical functional regions in the brain and neurons within them plays a significant role. And it would be interesting to know how this range is determined for information processing and shaping neural circuitry among the neurons. The Ising model is a simple physical model that could explain the ferromagnetic behaviour with domain coarsening of the two-spin lattice system through inter-molecular interactions imprinted by some range and concurrent computation of the Hamiltonian energy as a driving factor for the system evolution \cite{McCoy2014}. In our foregoing work, we proposed the \textit{neuron activity pattern or NAP model} which is based on the kinetic Ising model system \cite{Gundh2015}. In that work, we established a critical interaction range that demarcates the long and short ranges of interaction. Our approach evaluates the role of various interaction ranges on the strength of coupling and the overall Hamiltonian dynamics, as compared to the Onsager result that solved for only nearest-neighbour interactions \citep{onsager1944}. It has been shown in \citep{Gundh2015} that each interaction/coupling range has their unique critical temperature for generating spin clusters at the near-critical regime of phase transition though, their overall dynamical behaviour has followed universal scaling laws. Our anticipation led to the conjecture that cortical neurons might follow a second-order phase transition near-criticality to generate patterns of active neurons in response to a signal. However, the emergence of these functional clusters and their organization are still an open question. We have termed these patterns of neuronal activity as functional cortical patterns or FCPs.\\

{\noindent} In this study, we have generated model-based networks on the basis of, firstly, different ranges of interaction (long, critical and short) and secondly, another important parameter i.e. Temperature to define below, near and above criticality regimes. We have taken this architecture, further, to characterize temporal correlations based network connectivity of FCPs in our model data and found its congruency with the task-specific brain functional connectivity, attained through the empirical data. Since the functional connectivity reflects statistical dependencies between spatially distributed neuronal groups. We have made a comparison among networks of simulated and empirical data based on their topological characterization. This study has modelled the emergence of functional connectivity in the brain and to prove its fidelity we have applied different approaches such as network theory attributes, community detection, functional cartography of nodes and multi-fractal analysis. We have used the EEG data of a specific visual task on healthy human subjects for comparison with our model data \citep{Begleiter1999}. We have addressed important features of functional cluster formation, their organization and properties at below, near and above criticality for each interaction range. The major concern of this work is, whether the FCPs generated through our neuron {activity} pattern model, could {mimic} the topological characteristics of task-specific functional brain connectivity.

\begin{figure}
\label{fig1}
\begin{center}
\includegraphics[height=20cm,width=16cm]{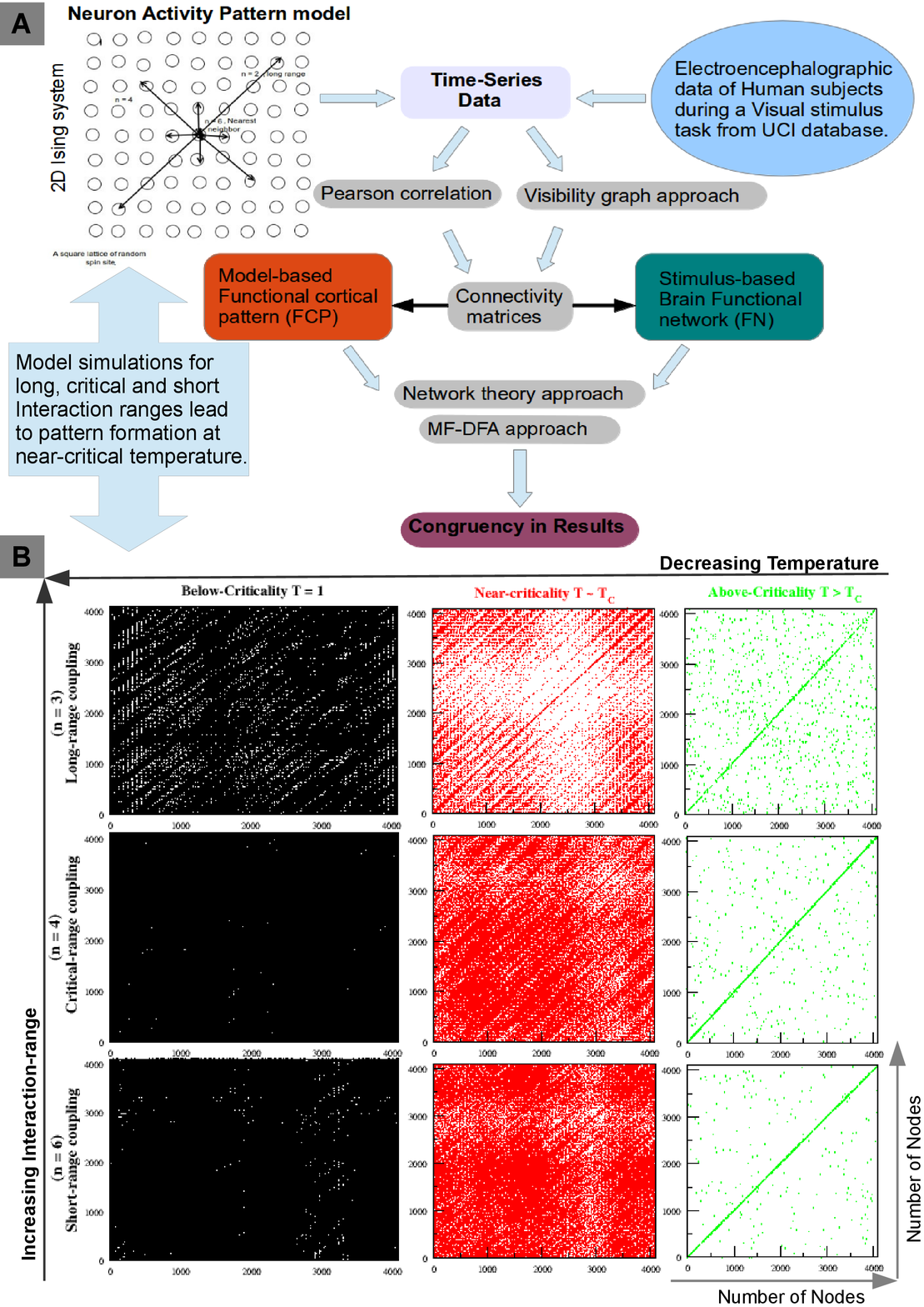}
\caption{\textbf{The Recurrence plots of model-based functional cortical patterns (FCPs):} \textbf{A)} Outline of the workflow used to characterize the model and empirical time-series data and to attain congruency in their functional activity patterns.
\textbf{B)} For Long ($n=3$), Critical ($n=4$) and Short ($n=6$) coupling
ranges at synaptic strengths below-critical ($T = 1$), near-critical
($T\sim T_{c}$) and above-critical ($T > T_{c}$) strengths ( as
shown in black, red and green respectively), in a network of 4096
nodes. \textbf{Note:- } There is the formation of communities of active
neurons at near-criticality for all the coupling range interactions.} 
\end{center}
\end{figure}

\vskip 0.5cm
\section{Results and Discussion}

We first need to delineate the details of the model simulation in the next sub-section, succeeded by the deliberation over results.

\subsection{Correlated neuron activity could optimize brain function}

{\noindent}The concept of self-organization rightly defends the dynamics of complex systems \citep{kauffman1993} and also aids in understanding brain functionality \citep{bassett2011}. Thinking of neural dynamics from a statistical mechanics window is a bit challenging but is ruling out since Ising model was used to explain the dynamical avalanches of neuron activity in the brain \citep{Chialvo2010,Eguiluz2005a}. In comparison with the Ising model a local neural network is not an equilibrium system as neurons constantly receives inputs from other neurons but the near-critical point of phase transition i.e. far from equilibrium could explain the dynamic equilibrium in complex systems. Also, the self-organizing dynamical systems expound that the brain works at a transition point when it receives the optimum signal strength \citep{Beggs2012}. Still, there is thirst for digging more into the complex dynamics of brain functionality. How brain does the transition towards a functional network that spans the cortical layers of brain, acquiring non-random synaptic connectivity for performing a cognitive task \citep{Song2005}.\\

{\noindent}This research work has used the NAP model that was introduced in our previous work \citep{Gundh2015}, therefore, the details regarding the Hamiltonian dynamics of the Ising model and its numerical simulation using the Monte Carlo Glauber kinetics with non-conserved magnetization has been briefly discussed here. The Ising model is one of the primitive models known to study phase ordering dynamics in a random distribution of spins under the influence of a critical parameter i.e. temperature (T). It had been used profoundly to explain the functional neuronal activity across functional brain networks \citep{Fraiman2009,goodarzinick2018}. This makes us strongly presume that phase transition state is the critical state of brain functional dynamics at which cognitive actions take place. We have considered random allocation of two-spin configuration in a 2D-square lattice with periodic boundary condition in order to avoid the edge effect. We contemplated the cortical neuronal circuits coarse-grained into the firing and non-firing states of neurons (or group of neurons) represented by the two states of a spin {s}, {s} $= +1$ for firing and {s} $=-1$ for rest or non-firing neurons. The total energy of the system representing various internal fluctuations and diffusive processes in neurons that result in their activity and influences the activity of surrounding neurons is given by the following Hamiltonian to model long as well as short-range interactions, 
\begin{eqnarray}
\label{2D}
H\:=-{\displaystyle {\displaystyle \sum_{<ij>}{\textstyle J(r_{ij},n)s_{i}s_{j}},}}\;\quad\;\:\;s_{i}=\pm1\,\forall\,i,
\end{eqnarray}
Where $s_{i}$ and $s_{j}$ denotes the activity of neuron at site \textit{i} and a neighbouring site \textit{j} and \textit{J} represents the coupling strength among neurons defined as a function of the distance between two $i$th and $j$th neurons given by, $r_{ij}=|r_i-r_j|$ which is  confined by $n$, the coupling range in the system. Then one can model $J$ by power-law behaviour given by, $J(r_{ij},n) =\frac{J}{r_{ij}^{n}}$ \citep{Cannas1996}, where, $d$ is the dimension of the system, and $n$ can be expressed as $n=d+\sigma=2+\sigma$ \citep{Fisher1972,Blanchard2013} which can qualify for \textit{short-range} for $\sigma>2$, whereas for \textit{long-range} for $0<\sigma<2$ \citep{Picco2012,gruneberg2004}. Further, $J> 0$ always (ferromagnetic case) and $n>0$ \citep{Cannas1996}. We have taken a 2D square lattice of spins with side $L$ and $N=L\times L$ being the total number of spin sites. To define coupling range we have assumed the centre of the 2D square as the reference origin to delimit the maximum distance for the spins to interact. The incircle of radius $r = L/2$ is the approximated area of the spin distribution which we have considered to calculate the potential energy, U, of the system. In order to capture the thermodynamical parameters of the system, one can define the following potential function $U(n)$ as a slowly decay potential function \citep{Chialvo2010,Cannas1996}, where $n =2 + \sigma$ for the 2D system, scaling $J\rightarrow J/N$ in the limit $N\rightarrow\infty$ and using Euler-McLaurin sum formula \citep{bruijn1981}, we get the potential energy function for each interaction range \textit{n} as follows, \citep{Gundh2015}
\begin{eqnarray}
\label{potential}
U\left(n\right)&=&\underset{N\rightarrow\infty}{\lim}\frac{1}{N}\sum_{<i,j>}^{N}\frac{J}{r_{ij}^{\sigma}}\approx \lim_{N\rightarrow\infty}J\int_{1}^{\sqrt{N}}drg(r)r^{3-n}\nonumber\\
&=&\lim_{N\rightarrow\infty}J\left[
\begin{matrix}
\frac{1}{2}\ln(N),~~~~~~~~~~~~~~~~for~~n=4~(critical)~~~~~~~~~~\\
\frac{1}{n-4}\left(1-N^{2-n/2}\right),~~for~~n>4~(short~range)~~~~~\\
\frac{1}{4-n}N^{2-n/2}~~~~~~~~~~~~for~~0<n<4~(long~range)~
\end{matrix}
\right.
\end{eqnarray}
where, g(r) is the pair distribution function such that $g(r) ~ 1$ for $r >> 1$. Critical parameters of the 2D system can be calculated from the force derived from the potential in equation (\ref{potential}), at the critical point where singularity arises in the solution of the system \citep{Hiley1965}. Following this technique one can estimate critical temperature from the numerically derived closed form approximated equation \citep{Hiley1965} given below,
\begin{eqnarray}
\label{closed}
\frac{U(n)}{k_BT_C}=1+\frac{f_2}{U(n)^2}+O(U^{-4})
\end{eqnarray}
where, $f_2=\sum_{<i,j>}J(r_{ij},n)^2$, which can be calculated for the 2D system by following the procedure mentioned above i.e. scaling $J\rightarrow J/N$  in the limit $N\rightarrow\infty$. Now using Euler-McLaurin sum formula \citep{bruijn1981}, $f_2$ is given by,
\begin{eqnarray}
\label{f2}
f_2&=&\underset{N\rightarrow\infty}{\lim}\frac{1}{N}\sum_{<i,j>}^{N}\left[\frac{J}{r_{ij}^{\sigma}}\right]^2\approx \lim_{N\rightarrow\infty}J^2\int_{1}^{\sqrt{N}}drg(r)r^{5-2n}\nonumber\\
&=&\lim_{N\rightarrow\infty}\frac{J^2}{2}\left[
\begin{matrix}
\left[1-\frac{1}{N}\right],~~~~~~for~~n=4~(critical)~~~~~~~~~~\\
\frac{1-N^{3-n}}{n-3},~~~~~~for~~n>4~(short~range)~~~~~\\
\frac{N^{3-n}-1}{3-n}~~~~~~~~for~~0<n<4~(long~range)~~
\end{matrix}
\right.
\end{eqnarray}
Now using equations (\ref{closed}), (\ref{potential}) and (\ref{f2}) we could able to arrive at an expression for finite critical temperature given below,
\begin{eqnarray}
\label{TC}
T_c(n,N)&=&\frac{1}{k_B}\frac{1}{1+\frac{f_2}{U(n)^2}}\nonumber\\
&=&\frac{J}{k_B}\left[
\begin{matrix}
\lim_{N\rightarrow\infty}\frac{N[\ln(N)]^3}{2N+N[\ln(N)]^2-1},~~~~~~~~~~~~~~~~~~~~~~~~~~~for~~n=4~(Critical)~~~~~~~~~~\\
\frac{4(n-3)}{(n-2)^2(n-4)},~~~~~~~~~~~~~~~~~~~~~~~~~~~~~~~~~~~~~~~~~~~for~~n>4~(short~range)~~~~~\\
\lim_{N\rightarrow\infty}\frac{2}{4-n}
\frac{N^{3(4-n)/2}}{2N^{4-n}+(4-n)^2\left[\frac{N^{3-n}-1}{3-n}\right]}~~~~~~~~~~~~~for~~0<n<4~(long~range)~~
\end{matrix}
\right.
\end{eqnarray}
From the above equations  (\ref{TC}), we could able to understand that for fixed systems size $N$ both short-range and long-range interaction of neurons contributed to $J$ lowers in $T_c$ allowing to work the neurons at a modified phase. On the other hand, for fixed $n$, $T_c$ increases as $N$ increases both in critical and long-range regime. The quantities such as range of interaction $n$ (or coupling range) and coupling strength $J$ controls the order parameter i.e. temperature $T$ of a spin-lattice system. However, there are debates on the estimation of $T_c$ dependence on $J$ in 2D system. Even though the relation between $T_c$ and $J$ was obtained analytically in 2D system as $sinh(2J/k_BT_c)\approx 1$ \citep{onsager1944}, numerically calculated $J/k_BT_c$ using Monte carlo techniques in such systems within finite size scaling formalism was found to be varying in the range $J/k_BT_c\rightarrow [2.269-2.29]$ \citep{Binder1988}. The reason for variation in the critical parameter value could be due to finite size ($N$) of the system \citep{Binder1988}, and $T_c$ depends on size $N$ except for short-range interaction ($n>4$). The critical temperature $T_c$ can be defined in terms of $J/k_B$, however, through our model simulations of random spin lattices, we have got the float values of $T_c$ for various long and short interaction ranges (see Fig.1 of ref. \citep{Gundh2015}), signified by the sudden drop in the total magnetization (order parameter) of the system.\\

{\noindent}Our idea states that this modelling framework mimics the underlying dynamics among a population of neurons. This study intends to relate the terminology being used in two different yet relatable systems, to characterize the brain as a dynamic physical system. We have defined the temperature, analogously, as the global synaptic strength given to a system of neurons (or groups of neurons) that generates the synchronous simultaneous activation of neurons sharing common synaptic strength. Therefore, the critical temperature would mirror the optimal perturbation (stimulus) imparted to the system i.e. relayed and associated in the form of synaptic strength in the brain. We have taken a 2D $64\times 64$ lattice system of 4096 spin sites depicting either firing or non-firing neuron state, allocated randomly, and simulated for 500 monte carlo steps at a defined global synaptic strength. Our approach examines the inter-neuronal connections for long, short and critical coupling ranges at different \textit{global} synaptic strengths ($T$) as categorized into three stages, below criticality, near-criticality and above criticality. The below criticality is defined as the temperature below the phase transition temperature but above the mean-field temperature i.e. $0 < T < T_c$. We have considered $T=1\frac{J}{k_B}$ as below critical temperature for all coupling ranges. However, the near-critical transition temperature, $T_c$, is different for the long $(T_c\sim 2.9\frac{J}{k_B})$, critical $(T_c\sim 1.9\frac{J}{k_B})$ and short $(T_c\sim 1.4\frac{J}{k_B})$ ranged interactions, based on their respective dynamics (see Fig.1 of ref. \citep{Gundh2015}). The spin-flip kinetics within Glauber kinetics algorithm defines the spin acceptance probability and also assumes ergodicity, however, the spin is chosen randomly \citep{Binder1988,newman1999}. The system quenches from a random disordered to ordered lattice phase and lead to the formation of domains/patterns of functional activity \citep{Sporns2000}. The Monte Carlo simulations for a particular coupling range $n$ and global synaptic strength $T$ has generated the temporal activity data based on the state of each neuron. In order to substantiate the global system \textit{behaviour,} we did average over temporal data of 10 ensemble sets in each case. The averaged time-series data is further used to construct binary undirected graphs by defining inter-neuronal connections using the Pearson correlation coefficient for different coupling range and strengths (see Methods). Thus, our analysis of a 2D lattice of neurons with fixed position and randomly allocated activity at each site lead to the emergence of FCPs of synchronous activity at the near-critical temperature, the optimum synaptic strength. \\ 

\begin{figure}
\label{fig2}
\begin{center}
\includegraphics[height=20cm,width=16cm]{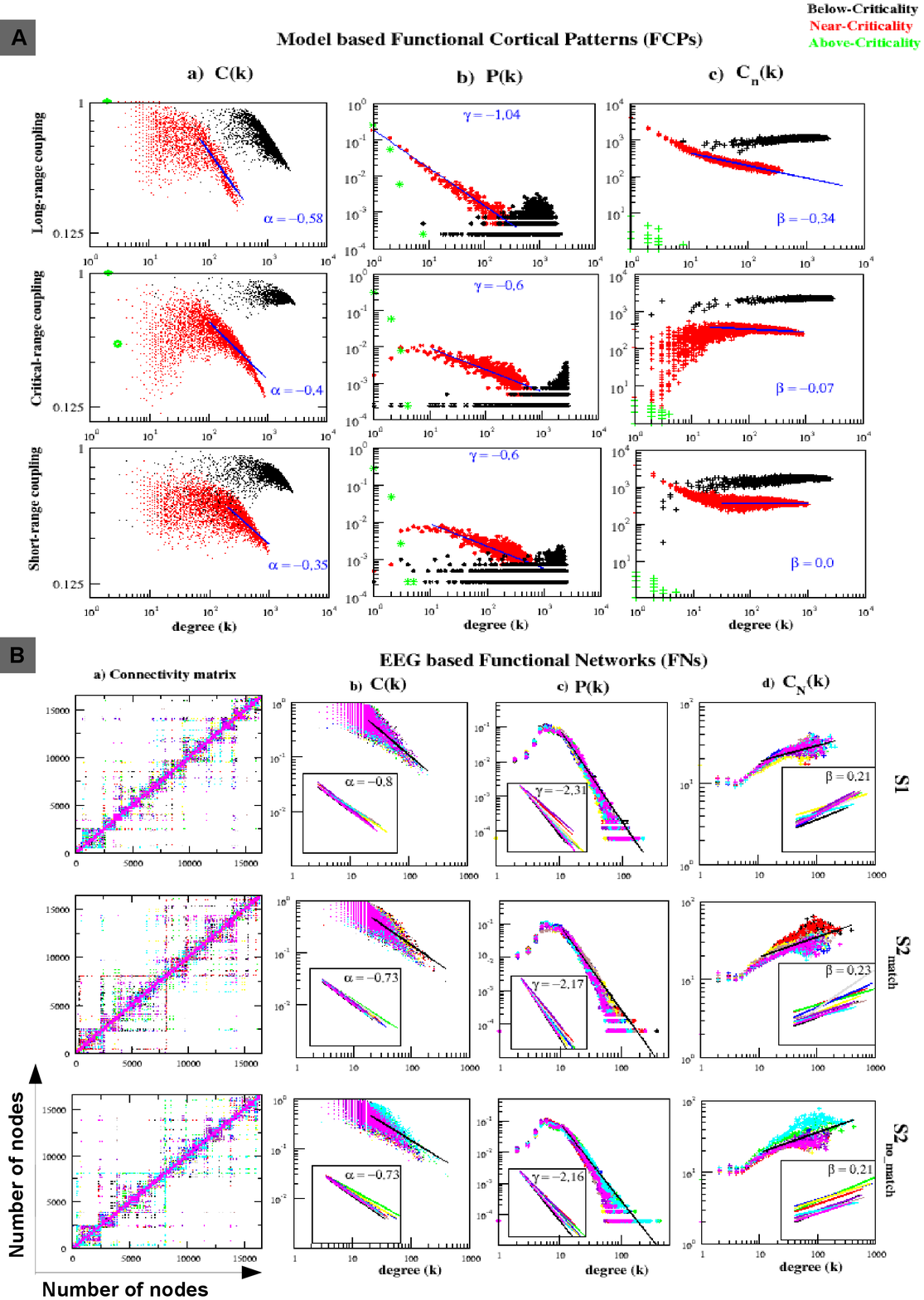}
\caption{\textbf{Basic network attributes of model-based FCPs and EEG-based
FNs:} \textbf{A)} FCPs for the three distinguished coupling ranges
at synaptic strengths below criticality (in black), near-criticality
(in red) and above criticality (in green). The columns exhibit a)
clustering coefficient, C(k), versus degree (k) plots with degree
exponent $\alpha$. b) The probability of degree distribution, P(k),
plots with degree exponent $\gamma$ c) The Neighborhood connectivity,
$C_{n}(k)$, plots with degree exponent $\beta$.
\textbf{B)} a) The connectivity trend of EEG data based FNs generated
for three visual stimulus conditions S1 (single image shown), S2\_match
(two similar images) and S2\_nomatch (two dissimilar images) of one
subject with 10 trials shown in different colours in each case. The
columns b), c) and d) illustrates the C(k), P(k) and $C_{n}(k)$
plots versus degree (k), respectively. The insider plots depict power-law fitting trend of 10 subjects with their average degree exponents.\textbf{Note -} In all the coupling ranges long, critical and short their
is transition shift at near-criticality (shown in red) for all the
three basic network attributes that characterize the hierarchical
and scale-free organization through the power-law behaviour of their
heavy-tailed distributions. The network measures are subsequently
in similar trend for near-critical FCPs and EEG based FNs, apart from
the obvious disparity among subjects.} 
\end{center}
\end{figure}

\subsection{Model-based functional cortical patterns}

{\noindent}The neuron activity patterns can be imprinted in a connectivity matrix based on the correlation in the activity of neurons. The temporal correlations of the neural activity has been translated in the form of binary connectivity matrices. We have applied different approaches for building the connection matrix out of the model and empirical time-series data as mentioned in the work flow-chart as shown in Fig.1A. We have tagged the model-based functional clusters of synchronous activity as FCPs. On the other hand, the visual task-based EEG time-series has been transformed into a functional brain network, FN, using the visibility graph approach \citep{Lacasa2008}. We have acquired significant congruence in the FCPs and FNs in their topological properties and functional characterization by applying the network theory and multi-fractal approaches \citep{Sporns2002a,Kantelhardt2002}. The model simulations at various interaction ranges and temperature strengths substantiate the emergence of functional patterns of activity only at the near-critical strength as depicted through the recurrence plots in Fig.1B. In this work, we have extended the interpretation of our model data and applied a fixed global synaptic strength which determines the intensity of overall coupling among the system of N neurons, being classified as below-criticality, near-criticality and above-criticality of the phase transition dynamics. The critical temperature $T_{c}$, which signifies the state of near-critical phase transition, is found to mimic with the optimum synaptic strength to coordinate short and long-range neurons in the brain for the generation of functional clusters of neurons at local and global levels.\\

{\noindent}The emergence of FCPs in our model is the result of self-organized criticality that formulates the functional pattern of neurons in the brain \citep{Kitzbichler2009}. These FCPs could mimic the functional networking that emerges in the brain cortex with task-specific activation of neurons. The recurrence plots in Fig.1B shows the emergence of FCPs being more prominent at near-critical temperature, $T_{c}$ for each coupling range. The long-ranged coupling at $T = 1$, is showing very less clustering and therefore it has many random connections not forming any ordered patterns or domains as reflected in recurrence plots. However, at near-critical synaptic strength ($T \sim T_{c}$), the transition to critical phase allows the formation of a number of functional patterns or domains. We could signify the near-critical transition as a state of attaining the optimal synaptic strength to construct synchronous functional patterns of neural activity and that could mimic the conscious state of a simple cognitive task in the brain \citep{Eguiluz2005a}. Further, at above-criticality ($T > T_{c}$) these patterns are violated due to enormous decline in the randomness of neuronal connections, causing break-down of neuronal self-organization. Also, the high synaptic strength could signify the abnormal state with loss of functional connectivity as observed in diseases such as Alzheimer's etc \citep{Stam2007}. The similar trend is followed by the critical and short coupling range (see in Fig.1B). However, the underlying dynamics is different for short and long-range interactions. The long-range coupling ($n=3$) has been speculated with fast dynamical clustering at near-critical temperature, {$T_{c}= 2.9\frac{J}{k_B}$} whereas, short-range coupling ($n=6$) has exhibited slow dynamics at {$T_{c} = 1.5\frac{J}{k_B}$} \citep{Gundh2015}. From equation (\ref{TC}), short-range $T_c$ $(n=6)$ is given by $T_c=\frac{3}{8}\frac{J}{k_B}$. Whereas, since $\lim_{n\rightarrow 3}\frac{N^{3-n}-1}{3-n}=\lim_{n\rightarrow 3}\frac{-(\ln(N))N^{3-n}}{-1}=\ln(N)$, the long-range $T_c$ $(n=3)$ is given by, 
\begin{eqnarray}
T_c(N)\approx\frac{J}{k_B}\frac{2N^{3/2}}{2N+\ln(N)}
\end{eqnarray}
This shows $T_c(N)$ depends on system size, $N$, {for long-range coupling} only. These changes in $T_c$ could suggest the enlargement of FCPs, generated with respect to short and long-range depending on $N$, with increasing synaptic strength signifying the local and global coupling over population of neurons \citep{Kitzbichler2009, Deco2012}. \\

\subsection{Network topology and characterization.}

{\noindent}The complex network theory approach, originated from graph theory, delineates the consanguineous activity of neurons in terms of quantitative measures \citep{Papo2014,Bullmore2009}. We have applied the network theory attributes here to characterize the functional networks generated through the neuro-physiological EEG data set and the simulated FCPs of the population of neurons.

\subsubsection{Properties of model-based FCPs:}

{\noindent}We have computed network theory attributes to study the topological properties of the FCPs generated (as shown in Fig.1), using our model approach. The network attributes calculated here, represents the topological information of the connected undirected networks generated for different coupling ranges of interaction. The sheer transition in the clustering coefficient distribution, $C(k)$, plots from below (black) to near-critical state (red) (Fig. 2A column $a$) apparently classifies the organization of a functional cluster in a hierarchical pattern as exhibited through the power-law scaling behaviour $C(k)\sim k^{-\alpha}$ as a function of degree $k$ with scaling exponent $\alpha$ for all the distinguished coupling ranges (long, critical and short). The data has been plotted on a logarithmic scale with non-linear curve fitting to expose the power-law scaling behaviour. The cases with an initial scattering of data points have shown that the power-law is exhibited only by their long heavy tail whereas, other cases have shown power-law scaling constituting whole or majority of data points. The power-law distribution has been confirmed by the algorithm given by Clausset et al. \citep{clauset2009}. However, at the above critical state (green) due to increased in the randomness of neuronal firing there are very fewer connections to form a functional pattern. Since $\alpha$ for long-range coupling provides larger value ($\alpha=0.58$) as compared to short-range coupling ($\alpha=0.35$), the long-range coupling at near-critical temperature probably enhances the active self-organization of neurons and local effects are allowed to participate in global phenomena. The probability of degree distribution, $P(k)$ again exhibits well defined fractal nature $P(k)\sim k^{-\gamma}$ at near-critical temperature with larger value of scaling parameter, $\gamma=1.04$, for long-range coupling and shows similar transition into a power-law function, thus, substantiate the presence of high degree nodes having less occurrence probability in the neural network. Similarly, the neighbourhood connectivity distribution also follows power-law nature, $C_{n}(k)\sim k^\beta$ with $k$, and near-critical temperature exhibits positive $\beta$ value for short-range and critical couplings indicating possibility of formation of rich-club, where some of the hubs hand in hand together to control the network. However, long-range coupling try to initiate neurons against this rich club formation (negative value in $\beta$). This assures that long-range coupling tries to keep hierarchical organization in the functional cortical pattern near-critical temperature phase. However, short and near-critical range couplings try to bring the network towards near scale-free features, where $C_n(k)$ becomes nearly independent of $k$ and $\beta$ becomes positive indicating significant importance of a group of hubs (rich club formation) respectively. Further, the behaviour at below critical temperature exhibits positive $\beta$ values (black colour data) for all types of coupling ranges (short, critical and long) which is a clear indication of rich club formation showing the significantly important role of hub groups in regulating FCPs. On the other hand, in the case of the above criticality regime, proper and clear behaviour of network properties in the FCPs are not exhibited for all coupling ranges.\\

\subsubsection{Properties of EEG based functional networks:}

{\noindent}The EEG time-series data of the whole brain while visualizing a particular single object (S1), two matching objects (S2 match) and two different objects (S2 nomatch) in human subjects has been taken from the UCI database. Since the networks or visibility graphs constructed from the EEG time-series data carry the properties embedded in the EEG data \citep{Lacasa2008}, characterization of these networks of various subjects may highlight how these brain functional networks work and organize themselves. Each node in the constructed networks corresponds to the time step mapped to the EEG time-series data which could be the resultant of the algebraic sum of all possible interacting neuronal signals (including all possible coupling ranges) at that time step. Hence, these nodes in the networks could be thought of as the representations of the interacting neurons at various cortical regions of the brain at various time steps collected by the EEG probes as signals. Further, these functional networks (FNs) thus constructed from the EEG data of the subjects, while doing a visual task, exhibit well-defined clusters of represented neurons with slight variation in their sizes and properties due to the different paradigms (S1, S2 match and S2 nomatch) (Fig. 2B column $a$). These FNs of the subjects shown to specific visual stimulus task declares the dynamics of the neuronal functional patterns specific to the visual cortex.\\

{\noindent}Now the connectivity plots and topological characterization of the EEG-based FNs show clear emergence of clusters/modules/domains of varied sizes in all subject cases (S1, S2 match and S2 nomatch) ( Fig.2B a). Similar to our theoretically obtained results (in Fig. 1), the behaviour of the topological properties of these networks (we presented the results only for 10 trials of a single subject and its also true for all trials of the subjects we have taken from the UCI database for analysis) follow power-law nature indicating the fractal nature of the system. More specifically, the properties of our neuron activity pattern model near-critical temperature is closely similar to the properties of FNs constructed from EEG data. Then, these topological parameters can be represented by,

\begin{eqnarray}
\bf{\Gamma(k)}\sim k^{F};~~\bf{\Gamma(k)}=\left[\begin{matrix}P(k)\\C(k)\\C_n(k) \end{matrix}\right],~~F=\left[\begin{matrix}\gamma\\ \alpha\\ \beta\end{matrix} \right]\rightarrow\left[\begin{matrix}-2.16-2.31\\-0.73-0.8\\0.21-0.23 \end{matrix}\right]
\end{eqnarray}

The positive value in exponent $\beta$ in $C_n$ indicates the assortativity property in the network (Methods) which enable significant hubs in the network to coordinate among them \citep{Colizza2006}. Further, the emerged modules/clusters follow hierarchical features and hence have a significant correlation among them (Methods). Because of these two reasons these FNs most likely exhibit \textit{rich-club} formation which is consistent with the experimental report on the brain \citep{teller2014}. This means that the rich-club formation of significant hubs in the hierarchically self-organized neuronal system may probably act to stabilize the system both at local and global levels.

{\noindent}Our interpretation theory professes as if a slow and weak signal (say at $T < T_{c}$) propagates among the cluster of neurons creating a pre-conscious state for neurons and then a perturbation (signal) would provide an optimal synaptic strength ($T_{c})$) that could drive the system towards a functional cognitive state. (This could give sense to the condition when there are many processes or thoughts that passes through our mind in an unconscious state and an external stimulus is required to attain the conscious state.) The $C(k)$ and $P(k)$ distributions have been found to be in accordance with the brain functional connectivity network attributes based on the experimental data as reported in \citep{Eguiluz2005a,Sporns2004}. The results shown in \citep{Eguiluz2005a} reflects the functional network attributes while listening to music and finger tapping. The complexity in functional brain networks has been classified as scale-free considering spatio-temporal neural activity through neuro-electrical and imaging techniques \citep{Beggs2003}. From this, we could affirm the validity of our model approach in characterization of the emerging FCPs formed in the brain while some specific cognitive task is done\citep{ThomasYeo2015}. Thus, the network representations of nodes at $T\,\simeq$$T_{c}$ for long, critical and short-range interactions assures the self-organization of nodes into an FCP which exhibits scale-invariant and hierarchical organization. This supports our understanding that the self-organization of the interacting neurons is maintained with the emergence of FCPs for disparate ranges at near-criticality \citep{Deco2012}. \\

\begin{figure}
\label{fig3}
\begin{center}
\includegraphics[height=12cm,width=16cm]{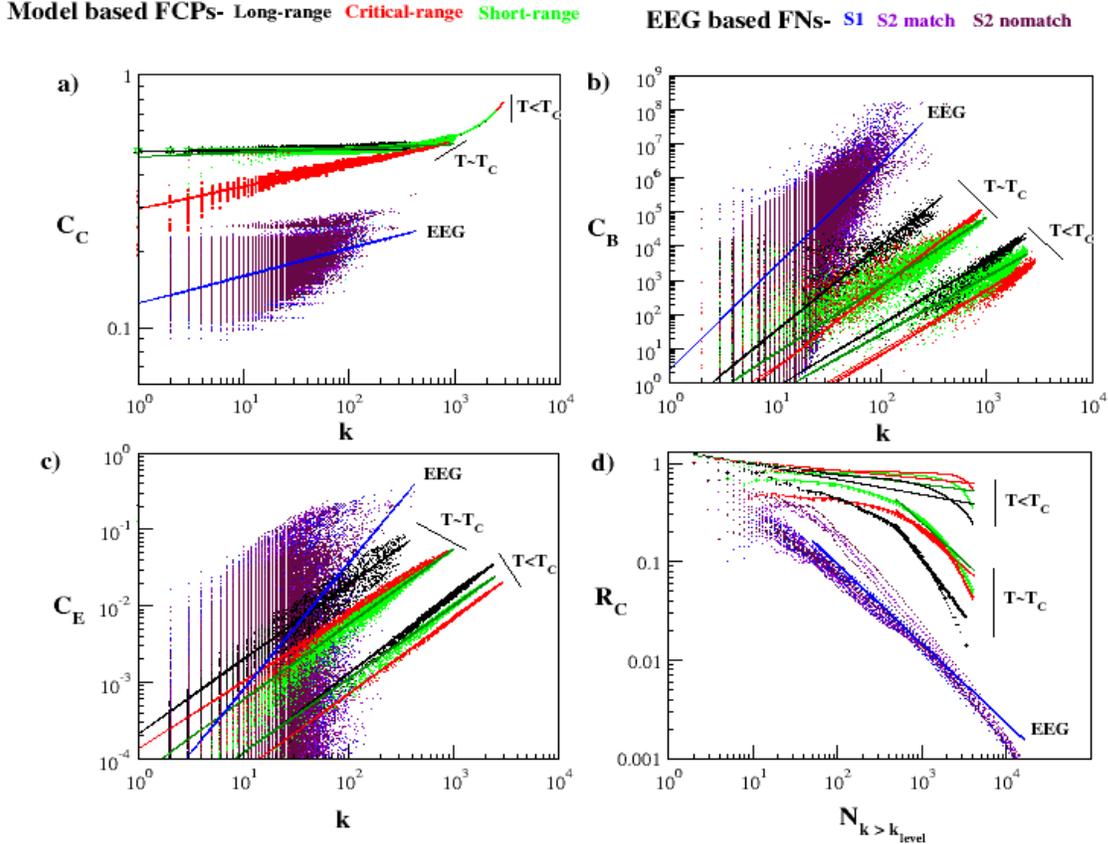}
\caption{\textbf{Comparative analysis of network attributes: }The model-based
FCPs have been analysed to mark the near-critical phase transition
in comparison with the EEG data based FNs. a) The closeness centrality,
$C_{C}(k)$, versus degree (k) plot with degree exponent $\delta$.
b) The betweenness centrality, $C_{B}(k),$ plot with degree exponent
$\lambda$. (C) The eigen-vector centrality, $C_{E}(k)$, plot with
degree exponent $\epsilon$. (D) The rich-club coefficient, $R_{C}(k)$,
is plotted against $N_{k>k_{level}}$i.e. number of nodes with degree
(k) greater than particular $k_{level}$. \textbf{Note}- The near-critical
FCPs have shown the right transition shift towards the FNs of a visual
task-based data. For a particular degree node the $C_{B}$ and $C_{E}$
value has increased to make its role more economical, though not much
change in $C_{C}$values observed. The $R_{C}(k)$ plot has shown
power-law trend with heavy tail distribution.} 
\end{center}
\end{figure}

\subsection{Centrality and rich-club measures: hierarchical to scale-free transition}

{\noindent}The roles and characterization of the significant hubs in the brain networks, in which emerged modules and sparsely distributed hubs work in a self-organized manner, can be well studied and compare using centrality measurements and rich-club analysis of the networks constructed from both neuron pattern activity model and EEG data. The centrality measures, as shown in Fig.3, characterizes the FCPs, at below and near-criticality for all coupling ranges and does a fair comparison with the FNs generated through the EEG data of a visual stimulus. Since the closeness centrality measurement $C_C$ of a node is estimated by the inverse of average distances of the node with the rest of the other remaining nodes \citep{Freeman1978}, a large value of $C_C$, which is quantified by small average path distance, provides significant importance of the node in regulating the network. The $C_C$ as a function of $k$ follows power-law nature, $C_C(k)\sim k^{\delta}$, both in proposed model network and network constructed from EEG data except the value $C_C$ in networks of EEG data is comparatively smaller than that of the model network (Fig. 3 (a)). The behaviour of $C_C$ for model network at near-critical temperature, $T\sim T_c$ for critical interaction range is closely similar to that of the network from EEG data, where, $C_C$ increases with $k$ indicating a significant role of high degree hubs in regulating brain network in terms of fast information processing of the neurons may be by coordinating the high degree nodes in all the modules of the networks. On the other hand, short and long-range correlation of the neurons again trigger more importance of the hubs moving towards the scale-free phase where hubs are more important than modules in controlling the networks. For higher degree nodes, the value is less at $T\,\simeq$$T_{c}$, this confirms the rareness of hubs in the network.\\

{\noindent}Betweenness centrality of the networks, which is another centrality measure from the number of paths passing through each node from the rest of the nodes in the network, quantify the volume of traffic diffusing through each node in these networks \citep{Freeman1978,borgatti2005}, and found that it follows power-law behaviour, $C_B\sim k^\lambda$ in both model and EEG data networks (Fig. 3 (b)). Since $C_B$ increases as $k$ increases, it indicates that high degree nodes have high traffic of information, and they play a crucial role to control and stabilize the network. In this case, the behaviour of the $C_B$ for all neuron coupling regimes (short, critical and long-range) are closely mimic with that of networks of EEG data in qualitative sense (indicated by nearly parallel fitting lines on the data points). This nature of $C_{B}(k)$, characterizes the cluster at $T\,\simeq$$T_{c}$ with higher value for high degree nodes and thus ensures greater significance of those nodes in fast information transmission across the network \citep{Bullmore2009}. Similarly, the properties of the eigen-vector centrality, which is the measure of influence of a node to any other nodes in the network that could induce long-term traffic risk \citep{bonacich1972}, show power-law nature, $C_E(k)\sim k^\epsilon$ to both for model and EEG data based networks indicating closely mimic behaviour (due to nearly parallel fitted lines). This eigenvector centrality, $C_{E}(k),$ further depicts the intensity of most prominent nodes in networks and found that network is highly communicative at $T\,\simeq$$T_{c}$ with low cost. Hence, high degree nodes have significant influencing capability to other nodes as well as more risk of attack to them. These high degree nodes could be generally target nodes of any brain disorder (disease, functional disorder etc).\\

{\noindent}Now the behaviour of all three centrality measurements ($C_C,C_B,C_E$) indicates the significantly important roles of high degree nodes in both model and EEG data based networks in terms of fast signal processing, regulation of information traffic, influencing other nodes in the network. This could be a clear signature of rich-club formation in these networks, indicating the significant role of the high degree hubs in regulating the networks but removal of these hubs will not cause network breakdown. Another thing depicts through our results are the presence of high degree nodes in long as well as short-range connections. This is in accordance with the experimental results done by \citep{Nigam2016}, where they found high degree neurons through direct recording from up to 500 neurons in slice cultures of mouse somatosensory cortex. In all centrality measurements of model based FCPs at $T\sim T_c$ and $T<T_c$, long-range neuronal coupling contribution is higher than the remaining critical and short-range interactions indicating a significantly more important role of long-range interaction of neurons in FCPs. One possible reason for more importance of long-range neuronal interaction in FCPs could be enhanced propagation of localized properties/perturbation, driven by short-range neuronal interaction, at the global level for better cross-talk to keep network stabilization in a self-organized manner.\\

{\noindent}The plot in Fig. 3d, quantifies the rich-club coefficient, $R_{C}$, as a function of $N_{k>k_{level}}$, which is the number of nodes with degree (k) greater than a particular $k_{level}$, where $k_{level}=1,...,max(k)$ (see Methods). This has encountered presence of hubs (high degree nodes) connecting module/communities patterned in a rich-club arrangement, to maximize communication at low-cost maintaining robustness of the network. The comparative analysis has shown that FCPs at below critical strength are far from the behaviour of the empirical data networks used here. However, at near-critical strength, FCPs leads to a shift towards the behaviour of FNs generated. The degree exponents of various network attributes calculated have been shown in Table-1 for all coupling range and strengths along with the EEG data. Apart from the visible differences in degree exponents of our model and EEG clusters, the trend followed by them in network attributes are quite similar.

\subsection{Modular organization: Is near-critical state active}

{\noindent}Modularity signifies the formation of community or modular structure in the network \citep{Newman2006}. The network centrality measures characterize the important nodes that determine the functional efficiency of the network. However, defining the role of a node based on its position within the module and participation among inter-linking modules is an effective way to portray complex networks \citep{Guimera2005}. In Fig.4A (a), we compare the community affiliated structure, ($C_{i}$), of long, critical and short-ranged coupling networks with the FNs of the EEG data. In column Fig. 4A (b), the within-module degree, $Z_{i}$, versus participation coefficient ($P_{i}$) plots relate the organization of influential nodes based on their intra- and inter-module links. The results show that the near-critical FCPs has marked good coherence with the EEG data. In Fig.4B, based on the categorization of nodes done in \citep{Guimera2005}, we have shown the transition in role and position of module nodes in the near-critical modular networks. The $P_{i}$ versus degree $k$ plots for the distinguished coupling ranges at below-critical and near-critical synaptic strengths has been shown in columns Fig. 4B (a) and (b) respectively. The nodes have been labelled as module hubs and non-hubs with $Z_{i}\geq2.5$ and $Z_{i}<2.5$ respectively. The further segregation depending on their $P_{i}$ values describes nodes as, $R1$ ultra-peripheral non-hubs ($P_{i}\leq0.05$) with all the edge connections in the same module; $R2$ peripheral non-hubs ($0.05<P_{i}\leq0.62$) with mostly intra-modular edges; $R3$ connector non-hubs ($0.62<P_{i}\leq0.80$) with many inter-modular edges; $R4$ kinless non-hubs ($P_{i}>0.80$) with homogeneous sharing of connections among modules; $R5$ provincial hubs ($P_{i}\leq0.30$) with most of the intra-modular connections; $R6$ connector hubs ($0.30<P_{i}\leq0.75$) with majority of inter-modular associations and $R7$ kinless hubs ($P_{i}>0.75$) with homogeneous associations among all the modules \citep{Guimera2005}. In our results of Fig.4B, the transition of temperature from below to near-critical regime leads to the occurrence of non-hub connector nodes, $R3$, (shown in green) in all coupling ranges. The early emergence of such hub nodes in long-range coupling could be due to its fast neuronal dynamics. Also, the near-critical FCPs has shown transition to the optimum hierarchical organization of connector hub nodes, $R6$, and peripheral non-hub nodes, $R2$, with respect to the degree for all coupling ranges, validating its functional modularity. The FNs of the EEG data (shown in violet), being a complex functional brain network, matches favourably with our model FCPs at the near-criticality regime. Further in Fig. 5, we have analyzed the probability of the degree distribution of nodes under a particular community, as constructed through the community affiliation Louvain method \citep{Blondel2008}. The modules generated through networks of below and near-critical FCPs has been shown in Fig. 5 (a) and (b), in respective different color for each module. For comparative analysis, the communities underlying the EEG data based networks have been shown in column (c) with quite a similar trend among various trials (shown in respective different colors) of a single subject. The below to near-critical transition signifies an increase in the number of communities. Also, the near-critical communities at short-range organize themselves towards small-world and tend to become hierarchical and scale-free at long-range as depicted through their degree distributions. This could be a signature to differentiate local and global coupling mechanisms in the functional brain networks which sounds helpful to tackle specificity of brain disorders. 

\begin{figure}
\label{fig4}
\begin{center}
\includegraphics[height=20cm,width=16cm]{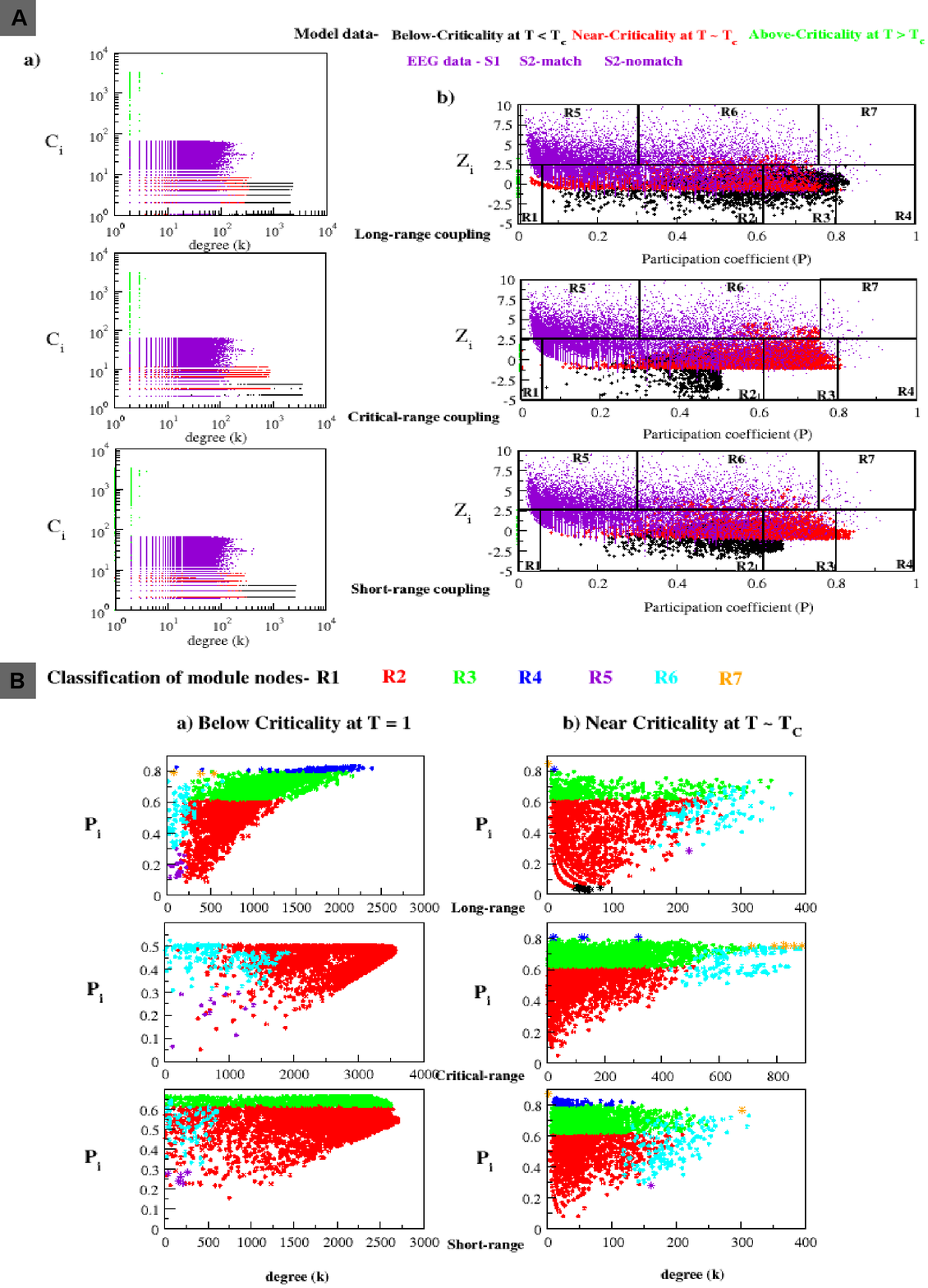}
\caption{\textbf{Community structure in networks of the model and EEG data: A)}
The model-based FCPs for the long, critical and short-ranged coupling
has been compared with the EEG data based FNs (shown in violet). a)
The community affiliation index, ($C_{i}$), versus degree (k) plots.
b) The within-module degree, $Z_{i}$, versus participation coefficient,
$P_{i}$, plots exhibits presence of module-hubs based on the classification
scheme defined as R1- ultra peripheral non-hubs, R2- peripheral non-hubs,
R3- connector non-hubs, R4- kinless non-hubs, R5- provincial hubs,
R6- connector hubs, R7- kinless hubs. \textbf{B) }The classification
of module nodes, based on the $Z_{i}$ and $P_{i}$ values, and their
organization depicted through the $P_{i}$ versus degree (k) plots
in the model generated FCPs for a) Below critical synaptic strength
at T = 1.0 and b) Near critical strength of their respective $T_{C}$.
The presence of various hub and non-hub nodes has been coded in respective
colors shown in the figure. \textbf{Note- }The congruency of the EEG
data (shown in violet) with the near-critical dynamics (shown in red)
claims the model robustness. There is emergence of R3 non-hub connector
nodes and R7 kinless hubs in the near- critical FCPs. Also, the organization
of R2 peripheral non-hub nodes and R6 connector hub nodes is showing
the right {hierarchy} with respect to degree upon transition to near
criticality.} 
\end{center}
\end{figure}

\begin{figure}
\label{fig5}
\begin{center}
\includegraphics[height=12cm,width=16cm]{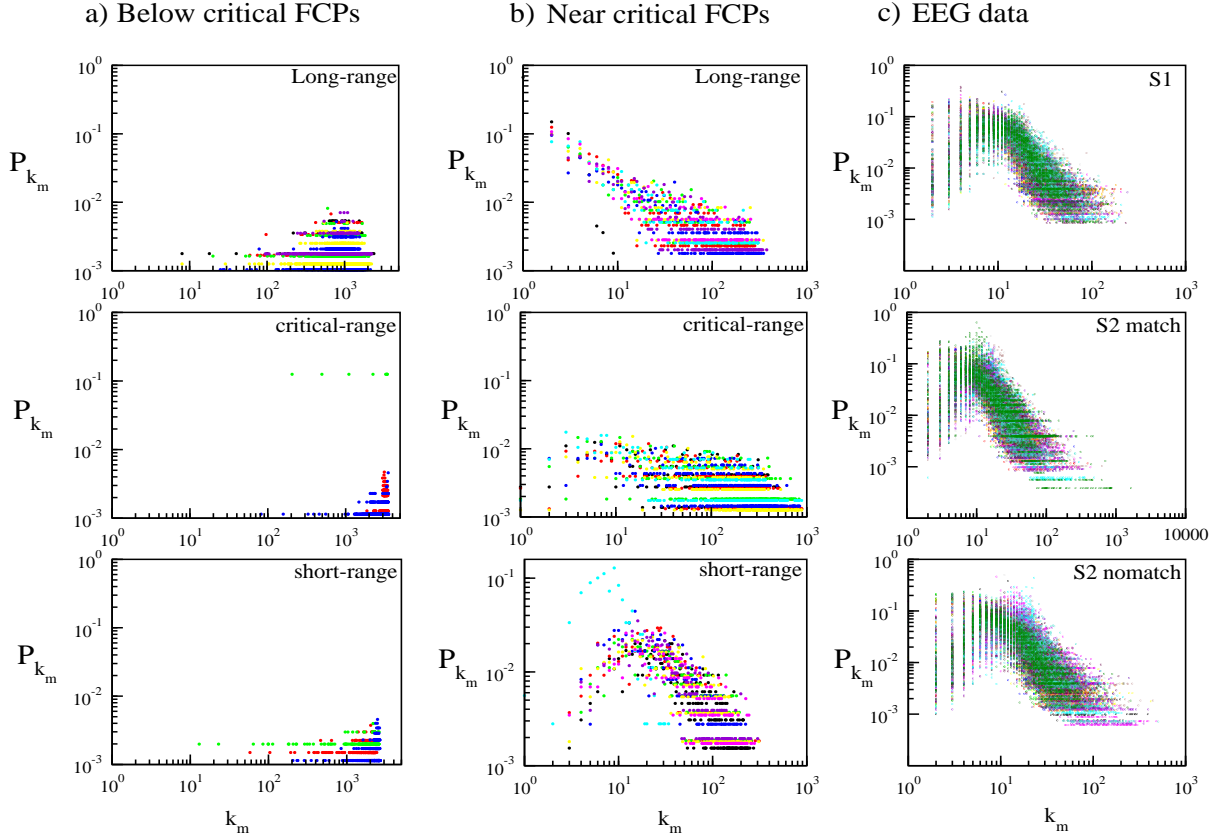}
\caption{\textbf{Analysis of Probability of degree-distributions in modules
of networks: }In a) and b) columns for long, critical and short-range
coupling we have shown probability of degree distribution, $P_{k_{m}}$,
of each node degree $k$ in module $m$ of the below and near-critical
FCPs, respectively. Here, different colors represent different communities.
c) The column shows $P_{k_{m}}$ versus $k_{m}$ plots of the communities
underlying the EEG data based networks for the three paradigms S1,
S2 match and S2 nomatch. Here, different colors represent communities
of various trials of a single subject. Note- The modules of near-critical
FCPs at short-range follows poisson distribution that transitions
to the scale-free organization at long-range coupling.} 
\end{center}
\end{figure}

\subsection{Multi-fractal signature due to complex brain organization}

{\noindent}Fractal characterization of a complex temporal signal has been done using the multi-fractal detrended fluctuation analysis (MFDFA) \citep{Kantelhardt2002, Song2005}. The power-law scaling behaviour exhibited by the networks at near-critical strength characterizes the approximately self-similar temporal pattern formation, which further announces the existence of a fractal dimension to describe the complex organization and role of short and long-range correlation of the components of the system. The comprehensive methodology on $MFDFA$ in \citep{Song2005, Ihlen2012} signifies the presence of multi-fractality in biological complex systems reflected in time-series data analysis of the system. We have studied the presence of multi-fractality in self-organizing FCPs at below and near-critical strengths and compared the same with the EEG data generated FNs.  In Fig.6 plot(a), the overall root mean square fluctuation function, $F_{q}$, for $q$ in range -5 to +5 is showing variation of fast and slow evolving fluctuations over segments of scale $s$ in the time-series, for each coupling range (long, short and critical) at $T<T_{c}$ and $T\simeq T_{c}$ and the EEG data as shown in respective colors. The scale of range 16 to 1024 has been used to calculate the function $F_{q}$ at each scale $s$ and is same for both the model and empirical EEG data. The slope of this scaling function $F_{q}$ determines the q-order Hurst exponent $H_{q}$. The $H_{q}$ is an important parameter to characterize multi-fractality as it captures the scaling behavior of small and large fluctuations in the time-series data, thus, its dependence on $q$ validates the multi-fractal nature. In Fig.6(b), the negative $q$ exhibits the scaling trend of small fluctuations and positive $q$ represents the scaling behaviour of large fluctuations in the data segments of a particular scale. The Fig6(b) has shown variation mainly in the scaling nature of small fluctuations at negative $q's$ and depicts an almost similar trend for large fluctuations at positive $q's$. The case of EEG data demonstrates the result of one subject each of S1, $S2_{match}$ and $S2_{nomatch}$ whereas the model data is the averaged result of 10 trials of time-series for all the three cases (long, short and critical). The $H_q$ of small and large fluctuations in the model data might get affected because of the averaging of data. However, we are more concerned about the scaling trend of the near-critical transition results being similar with the EEG data rather than the below criticality. The Fig6(c) represents the q-order mass exponent $t_{q}$ versus $q$, which exhibits the variation in mass exponent $t_q$ with respect to large ($+ve$ q) and small ($-ve$ q) fluctuations. The Fig6(d), partitioned into two subplots, further represents the generalized fractal dimension $D_q$ versus singularity exponent $h_q$ for below-criticality in the left subplot and near-criticality and EEG data in the right subplot. The mass exponent $t_{q}$, is a factor used to define the fractal dimension. 
In Fig6(d), the clear transition in the $h_{q}$ and $D_{q}$, from below to near-critical state, exhibits presence of multiple scaling exponents represented by the arc of the multi-fractal spectrum of long, critical and short coupling ranges that signifies an increase in complexity. The near-critical multi-fractal spectrum curves have shown high congruence with spectrum curves of the FNs of a visual task-based EEG data.
\begin{figure}
\label{fig6}
\begin{center}
\includegraphics[height=12cm,width=16cm]{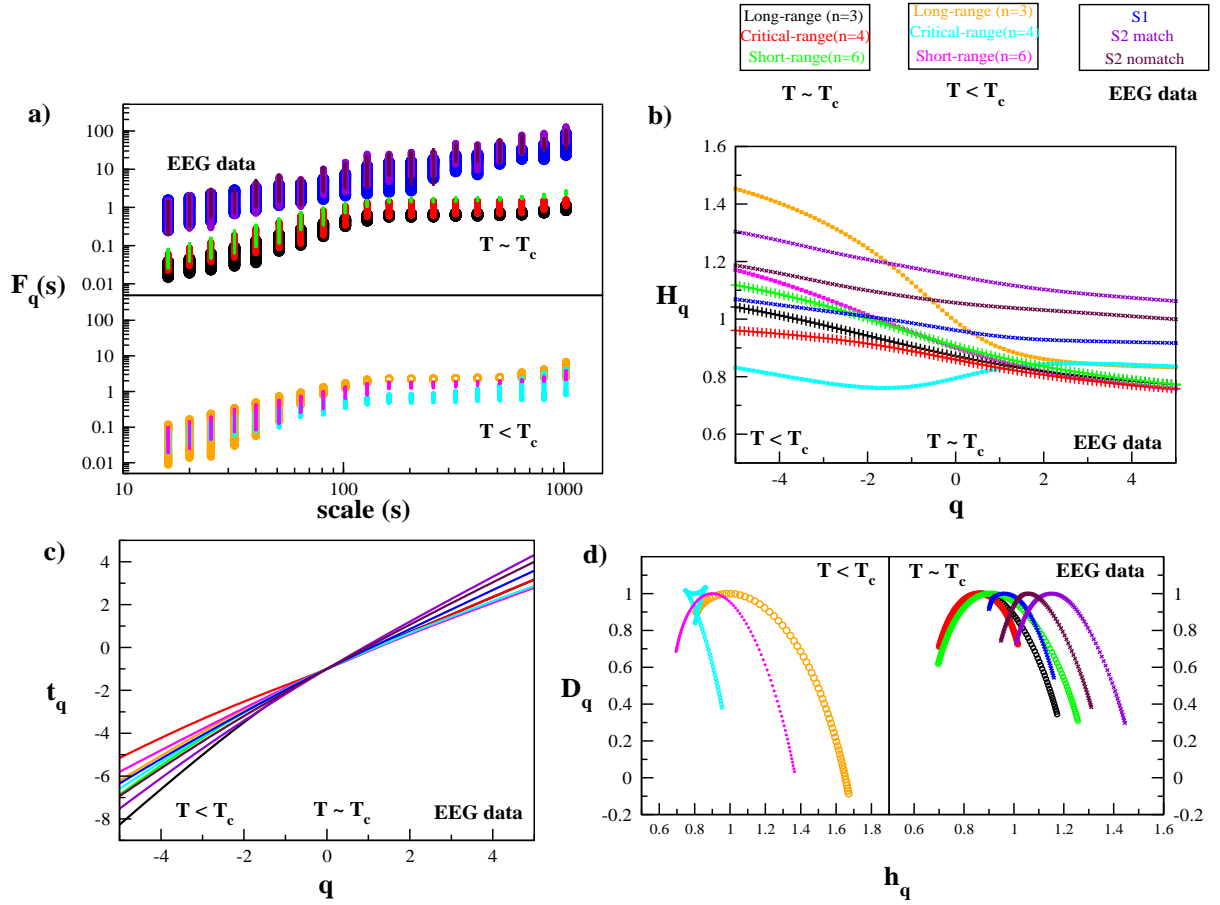}
\caption{\textbf{Multi-fractal detrended fluctuation analysis (MFDFA)
:} The technique has been applied to the model-based FCPs at below
and near-critical synaptic strengths and compared with the EEG-based
FNs to determine multi-fractal nature. a) The q-order RMS fluctuation
function $F_{q}$versus scale $s,$ quantifies the scaling trend in
fluctuations. b) The q-order Hurst exponent $H_{q}$, determines the
scale of the fluctuation function $F_{q}$. c) The q-order mass exponent
$t_{q}$ d) The q-order singularity exponent $h_{q}$ versus Dimension
$D_{q}$ exhibits the multifractal spectrum. \textbf{Note-} The transition
of FCPs at near-critical state is clear in the $D_{q}$ versus $h_{q}$
plot that exhibits a perfect arc of multifractal spectrum around the
range of exponents and complements with the FNs of a visual task-based
EEG data.} 
\end{center}
\end{figure}

\begin{table}
\begin{tabular}{|c|c|c|c|c|c|c|c|c|}
\hline 
\multicolumn{2}{|c|}{\textbf{Degree exponents:}} & \ensuremath{\alpha} & \ensuremath{\beta} & \ensuremath{\gamma} & \ensuremath{\delta} & \ensuremath{\epsilon} & \ensuremath{\lambda} & $\kappa$\tabularnewline
\hline 
\multirow{3}{*}{ FCPs ($T\langle T_{c}$)} & Long & -0.32 & -0.03 & 0.1 & 0.1 & 1.04 & 1.99 & -0.15\tabularnewline
\cline{2-9} 
 & Critical & -0.15 & -0.01 & 0.42 & 0.2 & 0.99 & 1.98 & -0.08\tabularnewline
\cline{2-9} 
 & Short & -0.34 & -0.07 & 0.14 & 0.26 & 0.96 & 2.0 & -0.10\tabularnewline
\hline 
\multirow{3}{*}{FCPs (T $\simeq$ $T_{c}$)} & Long & -0.58 & -0.34 & -1.04 & 0.0 & 0.98 & 2.16 & -1.11\tabularnewline
\cline{2-9} 
 & Critical & -0.4 & -0.07 & -0.6 & 0.08 & 0.88 & 1.97 & -0.72\tabularnewline
\cline{2-9} 
 & Short & -0.35 & 0.0 & -0.6 & 0.01 & 0.98 & 1.93 & -0.78\tabularnewline
\hline 
\multirow{3}{*}{EEG FNs} & S1 & -0.8 & 0.21 & -2.31 & 0.10 & 1.6 & 2.92 & -0.7\tabularnewline
\cline{2-9} 
 & S2 match & -0.73 & 0.23 & -2.1 & 0.10 & 1.8 & 2.34 & -0.78\tabularnewline
\cline{2-9} 
 & S2 nomatch & -0.73 & 0.21 & -2.1 & 0.11 & 1.56 & 2.15 & -0.87\tabularnewline
\hline 
\end{tabular}
\caption{The measure of degree exponents for various network theory attributes
(as mentioned in section 4.3).}
\end{table}

\section{Conclusion}

We have made an attempt to characterize cortical neuronal circuitry that emerges out in the brain during any cognitive assignment. In our model, the two-state dynamics has shown ordered pattern dynamics with the emergence of functional clusters, when it evolves at near-critical regime for all the interaction ranges. The emergent network connectivity at near-critical strength, that we call as FCPs, symbolizes the emergence of functional connectivity in the brain. We have done extensive topological characterization of the FCPs and compared with the task-specific functional brain network of human subjects. In this comprehensive work, we have validated our insights of brain functionality by characterizing properties of self-organized functional connectivity embedded in the time-series data of task-specific activity. We have analyzed combined effect of both coupling strength and coupling range as drivers defining the interactions in our model. 
Through graph theory measures, the effect of coupling range on the network of neural population depicts clear transition in the dynamic state of connectivity at the near-critical strength ($T\simeq T_{c}$). This illustrates the transition from a randomly connected network to a hierarchical network that owns scale-free characteristics with the presence of hubs as shown by the heavy tail distribution of degree distribution plots. The probability of degree distribution $P(k)$ and clustering coefficient $C(k)$ plots, clearly, explains the emergent complex dynamics of FCPs. The dynamics of the networks at $T\simeq T_{c}$ for long, critical and short-range coupling is found congruent with the emergent functional connectivity in the brain while performing a specific cognitive task. The self-organization in the brain lead to the emergence of FCPs in the form of hierarchically organized functional modules as reflected from the power-law nature of the rich-club organization and centrality measures of the networks constructed \citep{ThomasYeo2015, Meunier2010}. The presence of network centrality and rich-club behaviour, in similar trend with the EEG-based FNs, ensures the presence of coordinating hubs or dense modular cluster of neurons for effective communication in the model-based FCPs. This makes the functional pattern of neurons robust enough to maintain connectivity while wiring/rewiring among neurons as the dynamic pattern of connectivity among neurons affects the cost of information transmission \citep{DeDomenico2016}. The near-critical FCPs have shown the presence of $R6$, connector hubs, and $R7$, kinless hubs (in Fig.4B), with inter-modular connections to broadcast the information. The FCPs of neurons exhibit the hierarchical and scale-free organization with distinct community/pattern formation, where, coordination of the hubs via cross-talking communities in the network is one of the key factors of information processing both at local and global levels of neuronal organization. In our study, it is evident that this mechanism of FCPs organization is rectified by local and global perturbations in it, which can be propagated throughout the network with the help of various ranges of interaction (short, critical and long-range interaction). This manifests the fractal properties of these FCPs, which further exhibited multifractal spectrum formation at near-critical strength. Thus, the near-critical transition state marks the onset of cognitive state through self organizing dynamics at near-criticality. \\

{\noindent} Brain dynamics is known to function at near-criticality. There are recent studies that depicts the non-linearity of the functional brain dynamics at the near-critical regime \citep{Tagliazucchi2016, Cocchi2017, Ezaki2020}. Our results of network theory attributes, fractal analysis and functional cartography have confirmed the generation of a scale-free hierarchical network (FCPs) only at the near-critical phase transition. We have made the following insights focussing on the basis of generation of these FCPs. The dynamics of these FCP's could procreate diverse cognitive states in the brain. These cognitive states could be thought of as a bunch of meta-stable states in a system with near dynamic equilibrium. The near-critical regime could accommodate a set of dynamic functional states whose synchronous permutational amalgamation could lead its way to cognitive behaviours. Also, from our study of near-critical dynamics for long, critical and short-ranged functional patterns, we have interpreted that criticality is a range and not just a point. As we found that the long-ranged dynamics require high coupling strength to attain criticality than the short-ranged. This explains how the increase in synaptic strength due to multiple stimulus attempts would make more long-range connections to the complex functional pattern in the brain. The dynamicity of long-range connections defines the task-specific functional connectivity at the global level \citep{Park2013}. Also, such analytical {modelling} of functional networks of neurons based on inter-neuronal coupling range could further assist in the extensive assessment of brain disorders, characterized by aberrant functional connectivity. The research in \citep{goodarzinick2018} have analyzed 2D Ising model to exhibit topological characterization of functional networks of neuronal activity at criticality, being more robust against structural defects. However, in brain disorders, such as ASD irregularity in the structural connectivity has caused less intricate long-ranged and denser short-ranged functional connectivity \citep{Conti2017}. Also, the extent in reflection of structural connectivity changes on the topology of functional network cannot be determined till date and sounds debatable. Though, the research in \citep{goodarzinick2018} makes our belief more strong towards the critical phase transition state as the emerging functional state in the brain. We believe that our approach towards understanding the generation and characterization of functional patterns of neuronal activity in the brain cortex embarks us with new understanding on brain functional irregularities and also motivates modelling of functional connectivity in brain disorders.\\

{\noindent}There are obvious limitations that our model dynamics is binary and exhibits only firing/non-firing states of a neuron or group of neurons at a single lattice site. It overlooks the biochemical fluxes of a nerve cell and rather focuses on the global connectivity state. In our study, we have considered $64\times 64$ lattice system for our model and the empirical data is from 64 electrodes over the brain, this surely sounds a huge gap. However, brain activity is known to follow statistical fractal nature, which allows the characteristic properties to remain similar over the size \citep{Franca2018}. Also, we have addressed analytically that with an increase in system size the critical temperature will increase in the long-range regime, but this would not affect its near-critical dynamics. In spite of the huge complexity gap, we have got significant congruency in terms of the characteristic power law trends of the network topology between the task-specific functional brain networks and the FCPs at the near-critical regime only. This research has given us strong implication that the functional neuronal system supports far-from-equilibrium dynamics upon receiving a stimulus that leads to a second-order phase transition and self-organization is the key event that generates functional patterns in the brain cortex. This work gives us very useful insights towards the mechanistic view of the emerging brain functional connectivity, though there is still much more to dig and modelling through such simplistic model aid in understanding the concepts from a coarse lens that surely can be investigated to microscopic levels. \\

\section{Methodology}

\subsection{Network Construction:}

{\noindent}The functional connectivity of neuronal population in the brain could be examined through a binary connectivity matrix where '1' represents the connection between two nodes and '0' is antithesis of '1' \citep{sporns2007}. We have computed the functional connectivity based on statistical consanguinity of neuronal time-series using Pearson-correlation coefficient, used to measure inter-neuronal correlations \citep{sporns2007}. The Pearson correlation coefficient $r_{xy}$, defines correlation between nodes $x$ and $y$ using the formula,

\begin{eqnarray}
r_{xy}=\frac{\sum(x_{i}-\bar{x})(y_{i}-\bar{y})}{\sqrt{\sum(x_{i}-\bar{x})^{2}}\sqrt{\sum(y_{i}-\bar{y})^{2}}}
\end{eqnarray}

Where, $x_{i}$ and $y_{i}$ represents temporal data of two spins $x$ and $y$ for the total number of monte carlo steps. The binary undirected matrix is transformed into graph $G(S,E)$ consists of nodes set S = \{$s_{i}\mid i\:\in1$, ..., $N$\} where $N$ is the total spin sites i.e. 4096 in our study. Then, based on the functional correlation of spins the delineated edge list is defined as E = $\{e_{ij}\mid(s_{i},s_{j})\in S\times S\}$. We generated the binary connectivity matrix by averaging the data of 10 time-series from our model for each coupling range and global synaptic strength (Temperature). We applied a common thresholding limit for defining connections among the spins (neurons). We calculated the average correlation value in each case and then did the final average to get a unique threshold i.e. 0.2 for our model. This adjacency matrix is used to construct undirected networks for long, short and critical coupling range (n) at different overall synaptic strengths (T) employing NetworkX module in python. As we are studying the emergent behavior of our model generated functional patterns, we also constructed the functional brain cortical networks using the visibility graph approach on the EEG time-series data.

\subsection{Visibility graph approach:}

{\noindent}The EEG data has been taken from the UCI database, where the subjects were shown to three types of visual stimulus S1 (single image shown), S2\_match (two similar images) and S2\_nomatch (two dissimilar images) and recording was done using 64 electrodes with the sampling frequency of 256Hz for 1 second in healthy human subjects \citep{Beggs2012}. We applied the visibility graph approach in the EEG time-series data to construct the functional brain networks (FNs) generated while doing the visual task \citep{Lacasa2008}. We took 10 trials for each type of stimulus. In this approach, each time state or neuronal state in the EEG time-series (or neuron activity pattern model based time-series) data, which is the resultant signal of the interacting neurons in the brain at that time state, is taken as node in the constructed network. The connections between ant two time or neuronal states with data value $n_{a}$ and $n_{b}$ at time point $t_{a}$ and $t_{b}$ respectively would be defined if the third time 
step $n_{c}(t_{c})$ satisfies the following condition,

\begin{eqnarray}
\frac{n_{b}-n_{c}}{t_{b}-t_{c}}>\frac{n_{b}-n_{a}}{t_{b}-t_{a}}
\end{eqnarray}
The extracted network would be connected, at least to its neighbors, undirected and invariant according to the algorithm. The characteristic properties of the time-series get delineated in the form of resultant network.

\subsection{Network theory attributes:}

{\noindent}Network theory is the much applied theoretical approach by researchers for characterizing and studying complex brain networks \citep{Sporns2002a}. We are providing some important attributes, which are used for our analysis, as given below.\\
{\noindent}\textbf{Degree distribution}\\
The probability of degree distribution to have degree $k$ in a network defined by $G(N,E)$, where $N$ and $E$ are sets of nodes and edges of the network, is given by, $P(k)=\frac{n_k}{N}$, where, $n_k$ and $N$ are number of nodes having $k$ degree and size of the network, respectively \citep{Newman2009}. {The degree distribution, $P(k)$, for the Erd$\ddot{o}$s-Re$\acute{n}$yi random network follows Poisson distribution and deviates from the random for small-world networks}, whereas, for scale-free and hierarchical networks it follows power-law, $P(k)\sim k^{-\gamma}$, where, $2<\gamma<3$ \citep{albert2002,barabasi2004}. \\ 
{\noindent}\textbf{Clustering co-efficient}\\
Clustering co-efficient, which characterize how strongly a node is connected to the rest of the nodes in the network, of an ith node in the network, can be estimated as the ratio of the number of its nearest neighborhood edges to the total possible number of edges of degree $k_i$, $C(k_i)=\frac{E_i}{^{k_i}C_2}$, where, $E_i$ and $k_i$ are the number of connected pairs of nearest-neighbors of ith node \citep{Newman2009}. $C(k)\rightarrow constant$ for scale-free, random and small-world networks, whereas, for hierarchical network, $C(k)\sim k^{-\alpha}$, with $\alpha\sim 1$ \citep{albert2002,barabasi2004}.\\
{\noindent}\textbf{Neighborhood connectivity}\\
It is characterized by the measure of mean connectivities of the nearest neighbors of each node in a network given by, $C_n(k)=\sum_{u}uP(u|k)$, where, $P(u|k)$ is the probability that a link belongs to a node with degree $k$ points to a node with connectivity $u$ \citep{maslov2002,pastor2001}. The hierarchical network follows, $C_n(k)\sim k^{-\beta}$, with $\beta\sim 0.5$ \citep{pastor2001}. However, if $C_n(k)\sim k^{\beta}$ (positive $\beta$), then the network exhibits assortative nature indicating the possibility of coordinating high degree hubs in the network.\\

{\noindent}\textbf{Betweenness centrality}\\
Betweenness centrality of a node $w$ is the measure of sharing amount that a node $i$ needs $w$ to reach $j$ via shortest path, and is given by, $C_B(w)=\sum_{i,j;i\ne j\ne w}=\frac{d_{ij}(w)}{d_{ij}}$, where, $d_{ij}(w)$ is the number of geodesic paths from node $i$ to $j$ passing through $w$, and $d_{ij}$ indicates number of geodesic paths from node $i$ to $j$ \citep{Freeman1978}. It characterizes the amount of information traffic diffusing from each node to every other node in the network \citep{borgatti2005}.\\
{\noindent}\textbf{Closeness centrality}\\
Closeness centrality of a node $u$ is characterized by the harmonic mean of the geodesic paths connecting $u$ and any other nodes in the network, $C_C(u)=\frac{n}{\sum_id_{ui}}$, where, $d_{ui}$ is the geodesic path length between nodes $u$ and $i$, and $n$ is total number of nodes connected to node $u$ in the network \citep{Canright2004}. It generally measures the rate of information flow in the network, where, larger the value of $C_C$ corresponds short path lengths, and hence fast information processing in the network, and vice versa \citep{borgatti2005}.\\
{\noindent}\textbf{Eigen-vector centrality}\\
Eigen-vector centrality of a node $w$ in a network, which indicates the power of spreading of a particular node in the network, can be expressed by, $C_E(w)=\frac{1}{\lambda}\sum_{i=nn(w)}u_i$, where, $nn(w)$ is the nearest neighbors of node $w$, $\lambda$ is the eigenvalue of eigenvector $v_i$ $Av_i=\lambda v_i$, $A$ is the adjacency matrix of the network \citep{Canright2004}. The principal eigenvector of A, which satisfy $Au_w=\lambda_{max}$, is taken to have positive eigenvector centrality score.\\
{\noindent}\textbf{Rich-club co-efficient}\\
Rich nodes correspond to nodes having large links in the network, and they tend to form tight subgraph among them, which is referred to as rich-club formation in the network \citep{zhou2004}. This rich-club phenomenon can be quantified by rich-club coefficient defined by, $R_C(k)=\frac{E_{>k}}{^{N_{>k}}C_2}$, where, $E_{>k}$ as the number of edges after removing nodes less than degree $k$ distributed among $N_{>k}$ nodes \citep{Colizza2006}. It characterizes many important properties of the network, namely, information traffic backbone, mixing properties of the network, etc. The rich-club organization analysis has been done to estimate the presence of connector hubs. The Brain connectivity toolbox has been used to compute the rich-club coefficient and network centrality measures \citep{sporns2007,Rubinov2010}.\\
{\noindent}\textbf{Participation index}\\
{\noindent}The participation coefficient ($P_{i}$) signifies the distribution of connections of a particular node i with respect to different communities \citep{Guimera2005} given by,
\begin{eqnarray}
P_{i}=1-\sum_{c=1}^{N}\left(\frac{k_{ic}}{k_{i}}\right)^{2},
\end{eqnarray}
where, $k_{ic}$ is the number of connections made by node i to nodes in module c and $k_{i}$ is the total degree of node i. The $P_{i}$ value determines the distributional uniformity of the neuronal connections, specified by the range 0-1. The escalating value signifies a more homogeneous allocation of links among all the modules.\\

{\noindent}The \textbf{within-module degree} $Z_{i}$ is another measure to quantify the role of a particular node i in the module $c_{i}$. High values of $Z_{i}$ indicate more intra-community connections than inter-community
and vice-versa \citep{Guimera2005}.
\begin{eqnarray}
Z_{i}=\frac{k_{i}-\bar{k_{c_{i}}}}{\sigma_{k_{c_{i}}}},
\end{eqnarray}
where, $k_{i}$ is the number of connections of the node i to other nodes in its module $c_{i}$. $\bar{k_{c_{i}}}$ is the average k over all nodes in the module $c_{i}$ and $\sigma_{k_{c_{i}}}$ is the standard deviation of k in $c_{i}$.\\

\noindent\textbf{Community detection \textbf{method}:}\\
{\noindent}The complex network structure can be forbidden into communities or modules, specified with less than expected number of connections among them. We have used the community Louvain method to design non-overlapping communities out of the network \citep{Blondel2008}. This method has outperformed other methods in terms of both modularity and computational time \citep{Blondel2008}. To create a significant division of a network the benefit function called modularity (Q) is defined as, 
\begin{eqnarray}
Q=\frac{1}{W}\sum_{i,j=1}^{N}[A_{ij}-B_{ij}]\delta_{c_{i},c_{j}},
\end{eqnarray}
The modularity Q, is maximized for good partitioning of the graph $G(S,E)$ with N as total nodes. The $A_{ij}$ and $B_{ij}$ defines the exact and expected number of connections between nodes i and j. $W=\underset{i,j}{\sum}A_{ij}$, $c_{i}$and $c_{j}$are the communities nodes i and j belongs to,$\delta_{c_{i},c_{j}}$equals to 1, when nodes i and j falls in same community and 0, if they do not.

\subsection{MFDFA analysis:}

Multi-fractal detrended fluctuation analysis (MFDFA) approach has been used to ensure the fractality of dynamic patterns generated over non-stationary temporal data in complex biological systems \citep{Kantelhardt2002, Eke2002}. The statistical fractals generated through physiological time-series data show self affinity in terms of different scaling with respect to direction \citep{Eke2002}. The presence of multiple scaling exponents can be examined in a time-series using the Matlab formulation of the MF-DFA method \citep{Ihlen2012}. Parameters characterizing multifractality are scaling function (F), Hurst exponent (H), mass exponent (t), singularity exponent (h) and Dimension (D) as explained in \citep{Ihlen2012}. For a time-series signal $x_{j}$ of finite length \textit{l} with random walk like structure, can be computed by the Root mean square (RMS) variation, $X_{i}=\sum_{j=1}^{i}(x_{j}-\left\langle x\right\rangle)$, where, $\left\langle x\right\rangle$ is the mean value of the signal, and $i = 1,2, ...,
l$. The signal X has been divided into $n_{s}=int(\frac{l}{s})$ non-overlapping segments of equal size \textit{s}. To avoid left-over short segments at the end, the counting has been done from both sides therefore $2n_{s}$ segments are taken into account. This defines the scale (\textit{s}) to estimate the local fluctuations in the time-series. Thus, the overall RMS, F, for multiple scales can be computed using the equation,

\begin{eqnarray}
F^{2}(s,v)=\frac{1}{s}\sum_{i=1}^{s}\left\{ X[(v-1)s+i]-x_{v}(i)\right\} ^{2}
\end{eqnarray}
where, $v$ = 1,2, ..., $n_{s}$ and $x_{v}(i)$ is the fitting trend in each segment $v$. The q-order RMS fluctuation function further determines the impact of scale (s) with large (+ q's) and small (-q's) fluctuations, as follows,

\begin{eqnarray}
F_{q}(s)=\left\{ \frac{1}{2n_{s}}\sum_{\nu=1}^{2n_s}[F^{2}(v,s)]^{\frac{q}{2}}\right\} ^{\frac{1}{q}}
\end{eqnarray}
The q-dependent fluctuation function $F_{q}(s)$ for each scale (s) will quantify the scaling behaviour of the fluctuation function for each q,

\begin{eqnarray}
F_{q}(s)\thicksim s^{H_{q}}
\end{eqnarray}
where, $H_{q}$ is the generalized Hurst exponent, one of the parameters that characterize multi-fractality through small and large fluctuations ( negative and positive q's) in the time-series. The $H_{q}$ is related
to q-order mass exponent $t_{q}$ as follows,

\begin{eqnarray}
t_{q}=qH_{q}-1
\end{eqnarray}
from $t_{q}$ , the singularity exponent $h_{q}$ and Dimension $D_{q}$ is defined as,

\begin{eqnarray}
h_{q}=\frac{dt_{q}}{dq},\quad\quad D_{q}=qh_{q}-t_{q}
\end{eqnarray}
The plot of $D_{q}$ versus $h_{q}$ represents the multifractal spectrum of the time-series.

\section{Conflict of Interest Statement}
The authors declare no competing financial interests.

\section{Author Contributions}

R.K.B.S. and J.G. conceived the model. J.G. did the numerical experiment and prepared the figures of the numerical results. J.G. and R.K.B.S. analyzed and interpreted the results, and wrote the manuscript. All author read and approved the manuscript.

\section{Funding}
J.G. and R.K.B.S. are financially supported by Council of Scientific and Industrial Research through sanctioned project 25(0221)/13/EMR-II.

\section{Acknowledgments}
We acknowledge the funding agencies and Jawaharlal Nehru University. This manuscript has been released as a Pre-Print at [https://www.biorxiv.org/content/10.1101/569244v1] \citep{Jasleen}.

\section{Data Availability Statement}
The dataset analyzed for this study can be found in the [UCI machine learning repository: EEG database data set [https://archive.ics.uci.edu/ml/datasets/EEG+Database].


\end{document}